\documentclass[manuscript,screen]{acmart}
\usepackage{makecell}

\usepackage{graphics}
\usepackage{multirow}
\usepackage[T1]{fontenc}
\AtBeginDocument{%
  \providecommand\BibTeX{{%
    \normalfont B\kern-0.5em{\scshape i\kern-0.25em b}\kern-0.8em\TeX}}}

\setcopyright{acmcopyright}
\copyrightyear{2021}
\acmYear{2021}
\acmDOI{10.1145/1122445.1122456}
\acmConference[RecSys '21]{RecSys '21: 15th ACM Conference on Recommender Systems}{Sep 27--Oct 1, 2021}{Amsterdam, Netherlands}
\acmBooktitle{RecSys '21: 15th ACM Conference on Recommender Systems,
  Sep 27--Oct 1, 2021, Amsterdam, Netherlands}
\acmISBN{978-1-4503-XXXX-X/18/06}
\acmSubmissionID{1021}
\usepackage{wrapfig,lipsum,booktabs}
\usepackage{caption}
\usepackage{subcaption}

\begin{document}

\title{Privacy Preserving Inference of Personalized Content for Out of Matrix Users}

\author{Michael Sun}
\affiliation{%
  \institution{Stanford University}
  \streetaddress{450 Jane Stanford Way}
  \city{Stanford}
  \state{California}
  \country{USA}
  \postcode{94305–2004}
}
\email{msun415@stanford.edu}

\author{Tai Vu}
\affiliation{%
  \institution{Stanford University}
  \streetaddress{450 Jane Stanford Way}
  \city{Stanford}
  \state{California}
  \country{USA}
  \postcode{94305–2004}
}
\email{taivu@stanford.edu}

\author{Andrew Wang}
\affiliation{%
  \institution{Stanford University}
  \streetaddress{450 Jane Stanford Way}
  \city{Stanford}
  \state{California}
  \country{USA}
  \postcode{94305–2004}
}
\email{anwang@cs.stanford.edu}

\newcommand{\andrew}[1]{{{\textcolor{teal}{[Andrew: #1]}}}}
\newcommand{\tai}[1]{{{\textcolor{blue}{[Tai: #1]}}}}
\newcommand{\michael}[1]{{{\textcolor{magenta}{[Michael: #1]}}}}
\newcommand{\antonie}[1]{{{\textcolor{red}{[Antonie: #1]}}}}

\renewcommand{\shortauthors}{Sun, Vu, and Wang.}



\maketitle

\section{Introduction}

%The introduction explains the problem, why it's difficult, interesting, or important, how and why current methods succeed/fail at the problem, and explains the key ideas of your approach and results. Though an introduction covers similar material as an abstract, the introduction gives more space for motivation, detail, references to existing work, and to capture the reader's interest.

%Large pretrained models have delivered breakthroughs in areas across NLP, from Question Answering to Machine Translation, enabling a new generation of more content-aware consumer-facing recommender applications. 
\indent Recommendation systems in practice often rely on difficult-to-collect large-scale datasets in order to obtain a ``critical mass'' of user preference and content information for high-quality recommendations. With smaller datasets, the shortcomings of traditional recommendation systems begin to appear: sparsity of user preference matrices and item content lead to ineffective recommendations in data-driven systems \cite{ahmadian2019novel, singh2020scalability} particularly for cold start users and items. A large tail of niche application settings, such as the scientific and anime enthusiast communities, need recommender models catered to the needs of smaller, dynamic communities while keeping the options of privacy or anonymity.
%with a high degree of data sparsity, while being effective for cold start 
%particularly 1. a high percentage of first-time, i.e. ``cold start'', users and items, 2. privacy concerns within a smaller community and 3. rich semantic content, such as textual reviews and user-submitted item-to-item suggestions. 
%To build recommender systems for this long tail of niche communities, it is critical to make effective recommendations for ``cold-start'' users and items not seen during model fitting, while preserving the privacy of guest users and efficiently handling rich content with sparse training data. 

Existing approaches fail to provide cold start recommendations while maintaining the important needs of such communities.
Content-based cold start approaches \cite{bernardi2015continuous, lika2014facing} rely on user content information, which is invasive on user personal data, and does not work effectively in the setting of new users, for whom no preferences are given. %who have not provided a significant amount of their data to the system. 
Likewise, collaborative filtering \cite{su2006collaborative, chee2001rectree, 10.1145/963770.963774}, which provides recommendations based on preference-based similarity to other users or items, struggles with cold start users and items. While there are attempts to build more advanced recommendation engines using hybrid methods \cite{ctr, ctpr} and deep neural networks \cite{deepmusic, rnn, cdl}, research in alleviating cold start issues is still limited.
%we motivate the merits of a content-based approach with the problem of
%cold start, instances for which few or no interaction history is available (such as new users or added items).
We introduce \textbf{DeepNaniNet}, which uses a deep neural architecture to handle cold start with the option of learning rich content representations via a graph representation of the data, with edges linking users and items as well as edges linking items to items representing related show suggestions, enabling joint consideration of these data sources. %We further propose a graph neural network-based architecture to effectively model this rich graph structure to make recommendations.  
%In tandem with this graph representation, 
We introduce neural language embeddings derived from BERT to represent both textual reviews, which constitute the edges between users and optionally item-item recommendations, which constitute the edges between items. %We propose a pretraining and finetuning strategy within this recommendation framework to obtain embeddings tailored to the unique domain of anime reviews, given limited review data.
The model is trained with WMF following previous successes\cite{dropout}\cite{ctr}. Building on this framework, we introduce tools for generalization to new users that avoid profile mining: 1. a representation scheme of users by their ``content basket'', a set of a user's favorite items submitted to the service, with which to induce user representations, and 2. an autoencoder architecture to generalize to new users without requiring user metadata. 
%The inductive graph neural network approach, when combined with content basket representation and embedding dropout, enable effective recommendations in the cold start setting, both for new users and new items, while requiring minimal user personal data.
To our knowledge, DeepNaniNet is the first deep NLP driven anime recommendation engine. We further present AnimeULike, a new dataset of anime reviews consisting of 10000 animes and 13000 users, on which the system achieves competitive user recall on both warm and cold start animes and demonstrates a learnt ability to generalize to new anime on a thematic basis. On the benchmark of CiteULike, the system achieves equivalent performance to DropoutNet's SOTA cold start numbers and avoids any performance drop upon introduction of out-of-matrix users. In realistic real world settings where half or more users are either guests or have few suggestive interactions, DeepNaniNet proves to be the superior design. On AnimeULike Warm Start, DeepNaniNet demonstrates near 7x user recall@100 over WMF and 1.5x over DropoutNet, and further experiments indicate it is learning a rich user representation space in the process.

DeepNaniNet enables high-quality recommendations in the key stress areas for niche-community recommender systems. First, it can make effective cold start recommendations for both new users and items. Second, it does so via a privacy-preserving user experience and algorithm, which is crucial for both maintaining user anonymity and effectively serving guest users. Third, it is able to jointly learn to encode diverse data sources during training for qualitatively improved recommendation quality.

\section{Related Work}
%This section helps the reader understand the research context of your work, by providing an overview of existing work in the area.
\subsection{Traditional Methods}
%In recent decades, recommender systems have been an active area of machine learning research. 

The traditional class of solutions to building recommendation engines is collaborative filtering. Methods may model user representations via Bayesian inference of user preferences \cite{su2006collaborative}, unsupervised clustering of users with similar preferences \cite{chee2001rectree} or latent semantic analysis \cite{10.1145/963770.963774}. Dimensionality reduction via factorization of the preferences matrix has seen success in recommendation. Weighted Matrix Factorization (WMF) has been applied towards collaborative filtering \cite{hu2008collaborative}, and neural matrix factorizations have also been applied to recommendation \cite{xue2017deep}. Our method uses such factorization approaches to provide relevance scores to learn during training. Nonetheless, collaborative filtering approaches suffer from sparsity of user preference information. For example,  user-user and item-item collaborative filtering has no remedy for cold start users and items respectively when no preferences are available. Hybrid approaches to the problem of cold start in recommendation systems integrate in both user preferences and item content. CTR \cite{ctr} integrates latent embeddings for users and items with probabilistic topic modeling. CTPF \cite{ctpr} identifies latent topics and to form cross-topic recommendations. However, such methods introduce highly complex objective functions to incorporate additional content and preference terms. In addition, these models only handle cold start items without paying attention to cold start users.
%the ability to handle complex multimodal content information, such as textual descriptions and auxiliary social network or knowledge graph data

\subsubsection{Deep Learning Methods}

Meanwhile, many works have applied deep learning to developing both content and collaborative filtering-based recommendation engines, such as DeepMusic \cite{deepmusic} and RNNs \cite{rnn}, 
and CDL \cite{cdl}, which introduce various neural architectures to jointly learn content and preferences. Nevertheless, there have been limited attempts to address the cold start problem directly in the training procedure. DropoutNet \cite{dropout} demonstrated the SOTA result in generalizing to cold start users and items by employing dropout at training time to reconstruct user-item relevance scores. However, its shortcoming is in their sharp performance decline in situations where user content is not available. The present work combines the cold start advantages of DropoutNet while avoiding existing problems of privacy invasion posed by existing user content-based approaches \cite{recsurvey}\cite{socialtag}.

\subsection{Inductive Graph Learning}
Recently, general-purpose graph representation learning methods have been proposed for recommendation tasks. Pixie \cite{10.1145/3178876.3186183} uses biased random walks in the user-item interaction graph to generate recommendation scores, but does no learning, hence it cannot incorporate complex multimodal data end-to-end, and requires unwieldy hyperparameter tuning to design appropriate random walks for novel recommendation settings. PinSAGE \cite{ying2018graph} combines Graph Convolutional Networks \cite{kipf2017semi} with efficient inference and curriculum training for web-scale recommendation. However, PinSAGE does not explicitly address the cold start problem, which we aim to do in the present work. STAR-GCN \cite{zhang2019star} was proposed to handle the cold-start setting for both users and items. In the spirit of such inductive graph learning techniques, we leverage a graph neural network (GNN) architecture component to attain item feature representations that easily generalize to the cold start setting, while able to be trained end-to-end with other multimodal data sources. In particular, we make use of a GNN architecture based on GINEConv \cite{Hu*2020Strategies} to obtain node embeddings that encode textual information features, corresponding to reviews, along graph edges. In contrast to general-purpose graph representation learning methods, however, DeepNaniNet uses matrix factorization techniques which are more tailored to recommendation tasks, leading to superior performance as demonstrated in the experiments.

\section{Dataset}

\subsection{AnimeULike}
We propose a new dataset of anime reviews and ratings for recommendation. The dataset represents an impactful real-world recommendation setting which benefits a substantial niche community; it also carries key challenges of serving cold start animes and out-of-matrix users, thus serving as a benchmark for the generalizability of recommendation systems given sparse training data, with regard to privacy constraints. Here we describe the key properties and construction of the dataset.

\begin{table}
  \caption{Anime dataset specification with examples.
  The item-item anime graph has 27266 edges for 10000 animes.}
  \label{tab:exps}
  \resizebox{\columnwidth}{!}{%
  \begin{tabular}{llll}
    \toprule
    Section & Description & Specification & Example Entry \\
    \midrule
    \makecell[l]{User-Anime \\ Preferences} & \makecell[l]{Rating (1-10, 0 \\ for missing); \\ (optional) textual review} & $\mathbb{R}$ & \makecell[l]{8; "First of all, I have seen the original FMA and \\ although it was very popular and original, the \\ pacing and conclusion did not sit too well \\ with me\ldots"} \\
    %Brotherhood is meant to be \\ a remake of the original, this time sticking to the manga \\ all the way through, but there were people who thought \\ it would spoil the franchise. That myth should be dispelled, \\ as there's only one word to describe this series - EPIC."}  \\
    \midrule
    \makecell[l]{Anime-Anime \\ Suggestions} & \makecell[l]{Concatenation of all \\ written recommendations \\ made from anime1 \\ to anime2} & String & \makecell[l]{\textbf{"Hunter x Hunter (2011)} \\ Both are the best shounen in the world to me! \\ They start off with adventures of a brave main \\ character and go on into darker themes. \ldots \\%Love the way the character opens and \\ evolves throughout the story. 
    %Both series have an innovative and entertaining argument. %\\ The characters are so real; they are not the typical kick ass \\ character. For that reason, the strategies are the most \\ important at the moment of fight. \\ \ldots (in total from 96 users) 
    % \ldots\\
    \textbf{Shingeki no Kyojin}   \ldots"} \\
    \midrule
    Anime features & \makecell[l]{Numerical Features \\ (avg. rating, avg. rating \\ (rounded), \#ratings, \\ rank, \#members, \\ \#favorites)} & $\mathbb{R}^{6}$ & \makecell[l]{(9.18, 9, 1, 3, 2277948, 185915)}\\
    \midrule
    \makecell[l]{Anime Textual \\ Content} & \makecell[l]{Synopsis; Concatenation \\ of all written reviews} & String & \makecell[l]{"'In order for something to be obtained, something \\ of equal value must be lost.' Alchemy is bound by \\ this Law of Equivalent Exchange\ldots''}\\

    %—something the young brothers \\ Edward and Alphonse Elric only realize after attempting \\ human transmutation....";
    %\textit{concactenation of all reviews} \\ \textit{for anime from (user, anime) pairs}"}  \\
    \midrule
    \makecell[l]{User ratings} & Anime ratings by given user & $\mathbb{R}^{|U^{\text{train/val/test}}|}$ & $[9, 0, 0, \ldots, 8, \ldots]$  \\
     \midrule
    User content basket & Set of animes a user likes & Set<Item Ids> &  \makecell[l]{\{Fullmetal Alchemist: Brotherhood, Hunter x \\ Hunter, Code Geass\} (id mapped)}  \\
  \bottomrule
\end{tabular}%
}
\end{table}

\subsubsection*{Collection Pipeline}
We crawl a popular website of anime reviews and ratings, \url{MyAnimeList.net}, to obtain a rich graph of users and animes (``items'') as well as textual reviews and synopses. Out of respect for MyAnimeList.net's community, we avoid scraping user profiles directly. Instead, we crawl the listings of top rated animes to discover users, and additionally only save their id's to disk.

\subsubsection*{User-Item Reviews}
We obtain a graph of users and animes (items), connected by textual and numerical reviews. Our pipeline crawls through each of the site's top 10000 rated animes (as of Feb 2021). For each anime, we retrieve the numerical features and synopsis from its ``profile page'' and crawl its ``reviews page'', retrieving all available reviews, each of which comes with a written body, reviewer username, and rating (truncating at page 50 on the reviews to avoid spam). The list of users in the dataset is thus the set of all reviewers discovered this way. We send all users encountered during this phase to an additional sub-pipeline that crawls their past reviews for additional ratings. This way of extracting ratings from written reviews ensures ratings' integrity and retrieval of text that is directly responsible for predicting ratings. This collection step populates a preference matrix $R \in \mathbb{R}^{N\times M}$ whose rows are users, columns are items and entries are numerical ratings from a given user to the given item.

\subsubsection*{Item-Item Recommendations}
 To retrieve recommendations between a pair of animes, we crawl the ``userrecs" tab for each of the $10000$ animes' pages, representing written recommendations between two animes submitted by users, then concatenate all recommendations between each pair of animes into one body of text (made by $k$ recommenders --- we include $k$ as a metadata attribute \texttt{num\_recommenders}). Those with \texttt{num\_recommenders=1} are ignored, lest they be unsubstantiated or spam. This collection step results in a graph whose nodes are items and edges are annotated with textual recommendations between pairs of items.

\subsubsection*{Features}
Each anime page also comes with rich metadata and textual features. For each anime, we retrieved its MAL ID, numerical features (average rating, popularity, rank, members, favorites), and all reviews concatenated as one string. In total, we collected $27266$ anime pairs (each a long text body). Each (anime 1, anime 2) pair comes with all recommendations written on anime 1's page for anime 2 (an undirected edge). %, like \href{https://myanimelist.net/anime/5114/Fullmetal_Alchemist__Brotherhood/userrecs}{here}
Our entire pipeline is parallelized across both sub-pipelines and runs in ~20 mins. We also make the discovery and collection pipeline to be configurable. Our final user-item graph for training only contains an id for each anonymized user id's and the numerical rating (0-10) given. Our compiled dataset and web crawling codebase will be released fully upon deanonymization.

\subsubsection*{Preprocessing}

We first randomly split our dataset of $10000$ top-rated anime shows into training, validation and test 
with a $8:1:1$ ratio, resulting in 
$M^{\text{train}},M^{\text{val}}$, and $M^{\text{test}}$ animes in each split respectively. We %followed DropoutNet's paper by 
apply weighted matrix factorization (WMF) to approximate the training user-item rating matrix $R^{\text{train}} \in \mathbb{R}^{N\times M^{\text{train}}}$, as a matrix product $R_{u,v} \approx U_u^{\text{train}} {(V_v^{\text{train}})}^T$, where the rows of $U^{\text{train}}$ and $V^{\text{train}}$ are dense latent representations of the $N$ users and $M^{\text{train}}$ train set items, respectively.

\subsection{CiteULike}

In addition to our anime dataset, we tested our system on the CiteULike database \cite{bogers2008recommending}. This dataset contains 5551 users and 16980 articles, and each user has, on average, 37 articles. We demonstrate that DeepNaniNet achieves superior performance on widely-used benchmark CiteULike in the setting of out-of-matrix users, while preserving performance in the in-matrix case compared to the state of the art. To obtain item content $\phi_v^V$, we follow DropoutNet \cite{dropout} and run SVD on the TF-IDF top-8000 matrix, with 300 components, on the associated documents for CiteULike.

\section{DeepNaniNet: Recommendation Framework}
We propose a neural recommender system architecture, whose goal is to learn rich user and item representations that can predict user-item preferences given possibly incomplete, heterogeneous information (in the form of item and user content, or a partial view of user preferences). Within this system, we propose a ``user content basket'' technique for making accurate recommendations to users requiring minimal personal information, as well as an embedding dropout technique for improved out-of-matrix generalization. Together, these components form a robust system for generalizable recommendations on smaller, heterogeneous datasets.

\subsection*{Preliminaries}
We define the latent vectors obtained by WMF as $U_u \in \mathbb{R}^h$ for the latent vector of user $u$ and $V_v \in \mathbb{R}^h$ for the latent vector of item $v$. Let $V(u)$ be the set of all items user $u$ has interacted with $U(v)$ be the set of users an item $v$ has received ratings from. We presume that each item has an associated content vector denoted $\phi_v^V$. Depending on the dataset, each user may come with an associated content vector denoted $\phi_u^U$ too.

%Specifically, our AnimeULike dataset includes $13000+$ written recommendations. We define $G_1, G_2, G_3$ as GCNConv layers. \\

%We concat to $\phi_v^V$.

\subsection*{Architecture}
The architecture consists of a user encoder and an item encoder. Each encoder takes in a list of known user-item preferences $U_u$ and $V_v$ respectively, as well as a representation of any associated content (textual reviews or item-item suggestions), and outputs a latent embedding for the given user or item.

\subsubsection*{User and Item Encoder}
\begin{figure}[t]
    \centering
    \includegraphics[scale=0.3]{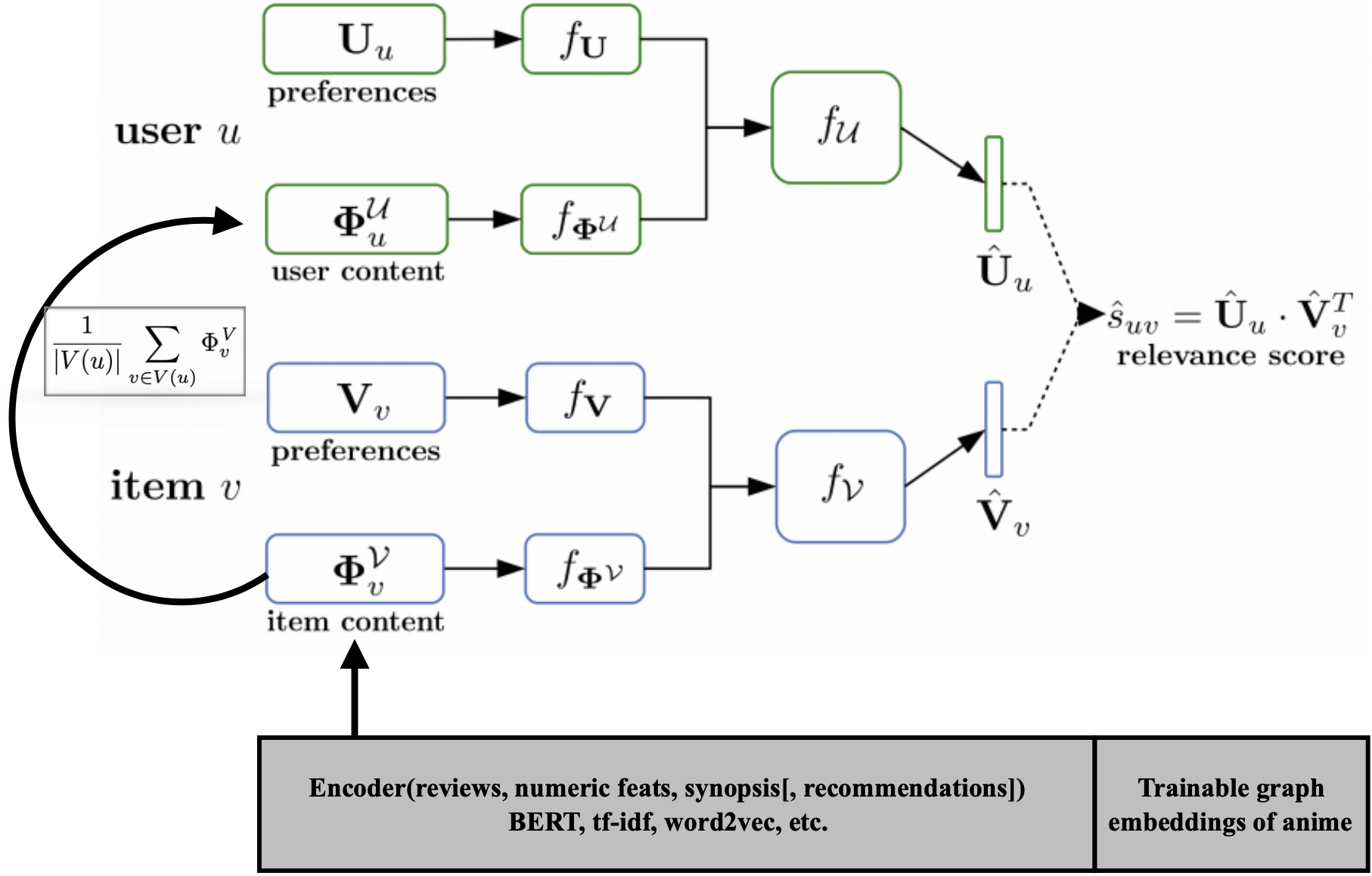}
    \caption{DeepNaniNet: Item content featurized by concatenating encoding with graph embedding \newline
    (Edited on top of DropoutNet paper's diagram \cite{dropout})} 
    \Description{}
    \label{ref:architecture}
\end{figure}

For each user $u$, the user encoder takes in its latent embeddings $U_u$ from the user-item preferences matrix, as well as its user content vector $\Phi_u^U$ (if provided), and outputs an encoder embedding $\hat{U}_u \in \mathbb{R}^r$, as shown in green in Fig \ref{ref:architecture}. The encoder proceeds as follows dropout. The user preferences vector $U_u$ is passed through a Dropout layer (rate $P_u$). The resulting user preferences vector after dropout is given by $\tilde{U}_u$ (see below). The user preferences vector after dropout $\tilde{U}_u$ is fed into a neural network layer $f_U\left(\tilde{U}_u\right) := \text{tanh}\left(W_U\tilde{U}_u\right)$ to produce an intermediate representation, and user content $\Phi_u^U$ is fed into a neural module $f_{\Phi^U}\left(\Phi_u^U\right) := \text{tanh}\left(W_U^\Phi\tilde{U}_u\right)$. The feature representations are concatenated and fed into a final layer $f_{\mathcal{U}}$: the output is given by:

\begin{equation} 
\hat{U}_u = f_{\mathcal{U}}\left(\left[f_U\left(U_u\right);f_{\phi^U}\left(\phi_u^U\right)\right]\right) := \text{tanh}\left(W_\mathcal{U}\left[f_U\left(U_u\right);f_{\phi^U}\left(\phi_u^U\right)\right]\right) \label{eq:useritemenc}
\end{equation}
where $[;]$ denotes concatenation. Similarly, for each item $v$, we pass $V_v$ and $\Phi_v^V$ to obtain $\hat{V}_v \in \mathbb{R}^r$, as shown in blue in Fig \ref{ref:architecture}. Following prior work \cite{dropout}, we standardize $U, V$ and apply batch normalization after $f_U, f_V, f_{\phi^U}, f_{\phi^V}$ with $500$ output units for $f_U, f_V$ and rank $r=200$ output units for $f_{\phi^U}, f_{\phi^V}$. %Depending on the content encoder's density, further hyperparameter search over number of layers and layer sizes. may prove worthwhile.

\subsubsection*{Content Encoder}
We experiment with choices for the content encoder $\phi^V$. When the content is textual (CiteULike documents or textual anime reviews), we take these functions to be either a BERT encoder or SVD on the tf-idf scores. Both methods take in a textual document and output a vector corresponding to that document, which we take as $\phi^V$. For BERT, we carry out a comparison of pretraining and finetuning strategies in the experiments to identify the most effective method for generalization to niche textual datasets. 

In particular, for the BERT encoders, we leveraged the pretrained weights of BERT in HuggingFace \cite{wolf2019huggingface} and finetuning it on our downstream task. In addition, we domain adapted our BERT models on all the texts in our dataset using masked language modeling \cite{devlin2018bert}, and then used these in-domain transferred models as our content encoders, with additional finetuning on the AnimeUReallyLike subdataset.

When the content is a graph $G = (V,E)$ (item-item suggestions for anime), with associated node feature matrix $X$, we compare the following graph neural networks (GNNs) to provide the content vectors $\Phi_v$:
\begin{itemize}
\item Graph Convolutional Network \cite{kipf2017semi}. The GNN takes in $(G,X)$ and a given item $v$ and outputs an embedding for the item, $f_{\phi^V}\left(\phi_v^V\right) := \hat{A}\text{ReLU}\left(\hat{A}XW_0\right)W_1$ for a two-layer GNN, where $W_0$ and $W_1$ are learnable parameters and $\hat{A}$ is the adjacency matrix after preprocessing according to \cite{kipf2017semi}.
\item GINE \cite{Hu*2020Strategies}. When there are additional features $V_{u,v}$ for each edge $(u,v) \in E$, we use GINE, which takes in $(G,X,V)$ and a given item $v$ and outputs an embedding for the item, which we take as $f_{\phi^V}\left(\phi_v^V\right)$. GINE performs multiple updates of the form:

\begin{equation}
 x_i^{k+1} \leftarrow \text{ReLU}\left(h\left(x_i^k + \sum_{j \in N(i)} \text{ReLU}\left(x_j + V_{j,i}\right)\right)\right)
\end{equation}
where $x_i^k$ is the embedding for the item after each layer $k$, $h$ is a neural network layer and $N(i)$ are the neighbors of node $i$ in $G$. We take $f_{\phi^V}\left(\phi_v^V\right) = x_v^K$ after a fixed number of GNN layers $K$.
%by representing each edge as an encoding of all recommendations written for the anime pair and the individual anime initialization as above.
\end{itemize}

Since availability of (item, item) edge content is contingent on the dataset, we concactenate $f_{\phi^V}\left(\phi_v^V\right)$ to the inputs to $f_{\phi^V}$ and later explore its ablation.

\subsubsection*{Relevance Score Prediction}
Given user $u$ and item $v$, the goal of the model is to predict the relevance score between the two. Given the output of the user and item encoders, we efficiently compute this score as $\hat{U}_u^T\hat{V}_v$, as shown in fig 1.

\subsection*{Our Solution for Cold Start}
For guest users (for whom $U_u$ is not available), we set $U_u=0$ and train our model to periodically "drop out" $U_u$ by relying on $\phi^U_u$ instead. Once a few interactions are collected, the user transform approximation $U_u\approx \frac{1}{|V(u)|}\sum_{v\in V(u)}V_v$ is taken instead - and vice-versa for cold start items. This is the approximation used by \cite{dropout} before each re-computation of WMF. We propose a new solution via an extension of the model's architecture instead.

\subsubsection*{User Content Basket}
One key insight is to model the user content as the average of its user's corresponding items:\\
\begin{equation}
    \phi_u^U= \mathbb{E}_{v\in V(u)}\phi^V_v\approx \frac{1}{|V(u)|}\phi_v^V.\label{eq:baskets}
\end{equation}This representation is advantageous from a modelling perspective because $\phi^U, \phi^V$ are in the same latent space, making the model more compatible with the autoencoder-inspired objective. This design decision also eliminates complications in user experience, with no privacy invasion from social profile mining, implications for user churn rate via soliciting for preferences or reduced service quality from exploration approaches that are difficult and expensive to implement. In particular, ``guest users'' who do not have prior information in the system can voluntarily submit a ``content basket'' $V(u)$ consisting of a small set of items that they prefer for inference. 

\subsubsection*{Learning to Generalize}
At training time, we explicitly train the model's ability to generalize to new items by performing one of three possible options:
\begin{itemize}
    \item Leave all embeddings as is.
    \item User Dropout: $\left(V_v,\phi_v^V\right)\rightarrow \left(\text{mask}(V_v),\phi_v^V\right)$ (with probability \texttt{user\_drop\_p}).
    \item Item Dropout: $\left(U_u,\phi_u^U\right)\rightarrow \left(\text{mask}(U_u),\phi_u^U\right)$ (with probability \texttt{item\_drop\_p}).

\end{itemize}

We note \cite{dropout} instead trains for cold start via training with both dropping out $U_u$ and the ``user transform'' $(U_u,\phi_u^U)\rightarrow \left(\frac{1}{|V(u)|}\cdot
    \sum_{v\in V(u)} V_v,\phi_u^U\right)$ and vice-versa ``item transform'' for out-of-matrix (unfactored) items. our solution has no need for this as inference is made solely with $\phi^U_u$ instead, with no loss of performance in reconstructing $U_u^T V_v$. We perform element-wise masking on (\texttt{user\_drop\_p} and \texttt{item\_drop\_p}) samples per minibatch to train for cold start (despite $V_v^{\text{val/test}}=0$ and/or $U_u^{\text{val/test}}$ in that setting). We observe improved cold start generalization this way with $\text{mask}=\text{Dropout}(\text{drop\_p})$, which contrasts with \cite{dropout}'s $\text{mask}=0$. Later, we apply $\text{mask}(U_u)=U_u+\text{Gaussian Noise}\left(\mu(U),\sigma(V)\right)$ to how these masking choices reveal DeepNaniNet's characteristics of a denoising autoencoder.

%We trained on this loss applying various choices encoder choices for $\phi^{V}$, which we term the \textbf{anime2vec} encoder.

%\subsection*{Model Architecture and Code}

%This section details your approach(es) to the problem. 
%For example, this is where you describe the architecture of your neural network(s), and any other key methods or algorithms.

\vspace{-1em}
\subsection*{Training}
To train the model, we minimize the loss:
 \begin{equation}
     L=\sum_{(u,v) \in S}\left(U_u^T V_v-\hat{U}_u ^T \hat{V}_v\right)^2 \label{eq:mseloss}
 \end{equation} via stochastic gradient descent, where $S$ is a set of training users and items. Intuitively, the learned embeddings should be able to reconstruct those learned by WMF, hence achieving the recommendation quality of WMF when full preferences are available. However, unlike WMF, the generalization ability of the neural framework will allow the model to make predictions about users not seen during training, as will be demonstrated in the experiments.
 
We note that the loss function is expensive to compute in practice, due to summing over $O(NM)$ terms. Hence, we approximate the loss via negative sampling, where $S$ consists of $k$ positive examples (pairs $(u,v)$ with nonzero preference matrix entry) and $5k$ negative examples (pairs with zero entry).

Throughout all experiments, we keep the same hyperparameters as in DropoutNet \cite{dropout} (e.g. dropout of 0.5), including the model architecture (single hidden units of 500 units), layer architecture (linear + batch normalization + tanh), and training data (as per trained WMF features and train/validation splits for CiteULike cold start), for consistency of evaluation. We follow the same practice of batching by users for each user sampling a fixed number of items, but differ in that we make sure each pass sees all true positive interactions. Unlike DropoutNet, we train without momentum for batch SGD, which we found to work better given this batching scheme. We sample random items at a ratio of 5:1 %(pos\_neg\_ratio) 
positives to negatives per epoch, ensuring each user sees a diverse set of item candidates. %We found that this procenables less sampled items per user and a smoother recall curve. 
This setup achieves superior performance on AnimeULike while reproducing reported results on DropoutNet. %the result of some early tuning of pos\_neg\_ratio differed just 0.009 from [1]'s reported SOTA of 0.636 and believe further optimizations can make up for the difference.\\% $U, V$ are passed through Dropout (with p=user\_drop\_p, item\_drop\_p) and $f_U, f_V$ respectively. \\

% The user and item feature vectors, $\phi_u^{U}, \phi_v^V$, are passed through $f_{\phi^U},f_{\phi^V}$ respectively. We then concatenated the latent and feature vector representations as $[f_U(U_u);f_{\phi^U}(\phi_u^U)]$ and $[f_V(V_v);f_{\phi^V}(\phi_v^V)]$ and pass them through $f_{\textit{U}}$ and $f_{\textit{V}}$ respectively to obtain $\hat{U}_u$ and $\hat{V}_v$.

\section{Experiments}
We run several experiments on both the anime and scientific article domains to understand generalization performance in terms of cold start and out-of-matrix users. Furthermore, we demonstrate robustness of the model to corruption rate. We run ablations to demonstrate the efficacy of our encoder components (GNN and domain-adapted BERT representations) in representing $\phi^V$ and $\phi^U$. Finally, we qualitatively analyze the method's recommendations in the case of AnimeULike and AnimeUReallyLike.

\subsection{Evaluation method}
%Describe the evaluation metric(s) you use, plus any other details necessary to understand your evaluation.
%Some projects will have clear metrics from prior work on given datasets, but we realize that other projects will define their own metrics.
%If you're defining your own metrics, be clear as to what you're hoping to measure with each evaluation method (whether quantitative or qualitative, automatic or human-defined!), and how it's defined.
We adopt user-oriented recall@K as our default quantitative metric (reported results set K=100 to stay consistent with prior work\cite{dropout}\cite{ctr}) and analyze qualitative patterns in a later section. We evaluate methods along two main scenarios.
\begin{itemize}
    \item In the \textit{warm start} setting, the system has observed the preferences $U_u$ for the user to retrieve recommendations for (resp. $V_v$ for items). Optionally, the system can incorporate user content $\phi^U_u$ (resp. $\phi^V_v$ for item content).
    \item In the \textit{cold start} setting, the system has observed no preferences for the user to retrieve recommendations for (i.e. $U_u = 0$, resp. $V_v = 0$ for items) and thus must rely solely on content $\phi^U_v$ (resp. $\phi^V_v)$. In cases where $\phi^U_u$ (or $\phi^V_v$) is not available, inference would not be possible. To address this, past approaches have proposed approximations of $U_u$ as the user's preferences become available. Instead, DeepNaniNet represents users via their content baskets (see Techniques for Improved Generalization), enabling effective inference in this setting from the very start.
\end{itemize}
%TODO: define warm start, cold start, in and out of matrix settings

\subsection{Preprocessing}
%Describe the dataset(s) you are using (provide references). If it's not already clear, make sure the associated task is clearly described.
%Being precise about the exact form of the input and output can be very useful for readers attempting to understand your work, especially if you've defined your own task.
For the anime dataset, we split the 10000 animes into train-val-test via a 8:1:1 split.  %themselves (8000-1000-1000), 
We apply WMF on the training preference matrix (of size $12767\times 8000$), excluding users who did not rate any items in the training set, to obtain a latent user matrix $U^{\text{train}}$ and latent item matrix $V^{\text{train}}$. For warm start, we include $1000$ validation animes (leaving out $1000$ for future experiments) and follow the same procedure as\cite{dropout}\cite{ctr} to refactorize a $12767\times 9000$ matrix and test on fold 1 (all cases nearly the same). We explicitly prevent corner cases where a user may be left with no animes to form his/her content basket. On average, the size of the user's content basket is $5.3$ in the cold start train fold, $0.6$ in the cold start val fold, $4.8$ on the warm start train fold, and $1.1$ on the warm start validation fold. 

We now define some important hyperparameters. For DropoutNet, we define \texttt{user\_transform\_p} as the rate by which we substitute user vectors via a ``user transform'' $U_u \approx \frac{1}{|V(u)|} \sum_{v\in V(u)}V_v$ (abbrev. ``UT'' in experiments), its proposed approach to handling out-of-matrix users. We drop out user embeddings in $U$ with probability \texttt{user\_drop\_p} and item embeddings in $V$ with probability \texttt{item\_drop\_p}. For DeepNaniNet, which utilizes the content basket representation, we take $U_u=0$ for out-of-matrix users at inference time (hence such users are represented entirely by $\phi^U_u$).% We believe this to be a reasonable upper bound to out-of-matrix users performance in the general case, as held out users would reduce performance and synthetic user samples could be more biased than the observed samples $U_1, U_2, \ldots, U_m$ from the $U$ distribution (though we leave this for future study).

\subsection{Experiments} 

\subsubsection{CiteULike Cold Start}
The user transform (UT) approximation used in DropoutNet, while resulting in performance gains for out-of-matrix users, causes a a significant performance drop for in-matrix users, hence cannot serve both classes of users simultaneously. This suggests that DropoutNet overfits to $U$, and approximations of the user representation $U_u$ (even by a good approximation since $|V(u)|\approx 30$ for CiteULike), poses a significant handicap. Poor out-of-matrix performance also has negative consequences for industrial models, necessitating that they be retrained entirely after each computation of $WMF$ as $U$ and $V$ shift due to evolving distribution of users and documents over time. 

\begin{wraptable}{l}{8cm}
\vspace{-1em}
\caption{On recommendation for CiteULike, DeepNaniNet gracefully handles both the in- and out-of-matrix cases, achieving comparable or favorable performance in both cases. In contrast, DropoutNet is not able to handle both cases simultaneously, even with the user transform technique for handling out-of-matrix users, due to the overfitting to $U$.}
  \label{tab:cite_coldstart_results_new}
%   \vspace{-1em}
  \begin{tabular}{p{2cm}p{1.5cm}p{1.5cm}p{1.5cm}}
    \toprule
     & In-matrix & 50\%-50\% & Out-of-matrix \\
    \midrule
   DN (no UT) & \textbf{62.1} & 59.1 & 56.7 \\
   DN (UT) & 7.7 & 30.9 & 55.4 \\
   \midrule
   DNN & 61.3 & \textbf{61.2} & \textbf{61.0} \\
  \bottomrule
\end{tabular}
\end{wraptable} 
In realistic real world settings where up to half of users could be guests, this limitation  implies the impracticality of serving guest or newly registered users (for whom there are no or few interactions) without either a) recomputing $WMF$ after sufficient exploration or b) finding more invasive ways to obtain better user representations. While DropoutNet holds onto a slight edge in the case of all in-matrix users (despite DeepNaniNet having access to $U$ all the same), we see this as a positive indication that DeepNaniNet is instead learning a rich latent space beyond merely information contained in $U$ or $V$.
\break \vspace{-1em}
\subsubsection{AnimeULike Warm Start}
We next compare anime recommendation performance in the warm start setting. For AnimeULike, we keep the same setup as before. We experimented with both TF-IDF and transferred BERT as the content encoder. Our base model experiments reveal transferred BERT as the better $\phi^V$ anime2vec encoder for \textit{both} DropoutNet and DeepNaniNet (whereas in the CiteULike domain, TF-IDF has the edge). Thus, to put the best foot forward for both models, we fix the domain-adapted BERT as the anime2vec encoder for all warm start AnimeULike experiments.

\begin{wraptable}[15]{l}{8cm}
\vspace{-1.5em}
\captionof{table}{AnimeULike: Warm Start. DeepNaniNet surpasses DropoutNet for both in-matrix and out-of-matrix users. In brackets are preliminary runs with tf-idf as the content encoder.}
\vspace{-1em}
\begin{tabular}{p{3cm}p{2cm}p{2cm}}
    \toprule
    Method & In-matrix & Out-of-matrix \\
    \midrule
    WMF & 9.4 & 8.1  \\
    Popularity sort & N/A & 2.4 \\
    Random guessing & N/A & 1.1 \\
    DN & 38.3[/37.6] &36.7[/36.4] \\
    DN (UT) & 37.5 &36.9\\
    DN (added GNN) & 42.0&39.2  \\
    DN (added GNN, UT)  & 41.0&39.1  \\
    %DN (GNN + NO EIP + NO UT) & & \\
    %DN (GNN + NO EIP + HALF UT) & & \\
    %DN (NO GNN + NO EIP + NO UT)  & & \\
  % DN (NO GNN + NO EIP + HALF UT) & & \\
    \midrule
    DNN (removed GNN) & \textbf{64.3}[/61.0]&\textbf{61.4}[/60.9]  \\
    DNN (V=0) & 63.3&61.1 \\
    DNN (full) & 64.2  & 59.7 \\
  \bottomrule
\end{tabular}
\end{wraptable} 
\vspace{0.5em} Our initial observation is the magnitude-fold improvement over the WMF baseline. In \cite{dropout}, only a 0.001 improvement is made over the WMF baseline on warm start. The poor metrics of WMF warm start recommendations seems consistent with users' experiences using our demo, with one user source commenting, "I typed in Flip Flappers and it just gave me the FMAB, AOT S3.5, and AOT S4 despite them being the exact opposite of what I'm looking for. Looks like it just picked the top 3 rated shows from MAL." As the \textit{popularity sort} baseline shows, correlation with popularity is low, so methods like WMF and CF that learn off of preference overlap between users perform poorly. The sparsity (~5 items/user for warm start) and stochasticity of user anime interactions produces much poorer approximations of $U$ and $V$ than as in CiteUlike, forcing the autoencoder to rely on content instead. This is substantiated by the fact DeepNaniNet jumps to \textit{over double} the performance of DropoutNet from the very start. We run each model the same number of updates until performance plateaus. While DropoutNet makes up for a little differential, it's clear by now using content baskets is the superior design. Visually, there's a large separation between the two families of models' performances. Adding GNN to DN helps some, but the performance bottleneck is most likely the expressiveness of $U$, hence leading us to conclude our performance owes itself to the content basket representations. This suggests our model is constructing a rich shared representation space between $(\phi^V,\phi^U)$ and $(V,U)$ from the start, exhibiting the ideal behavior of a denoising autoencoder and thus avoiding the discussed pitfalls of overfitting to $U$ or $V$. In fact, we observe swapping $V_v=0$ at evaluation time not only doesn't hurt performance but increases it marginally.\\

\begin{wraptable}{l}{5cm}
\vspace{-2em}
\captionof{table}{DeepNaniNet is robust to noise corruption of up to $\approx 70\%$ samples.}
  \begin{tabular}{p{2cm}p{2cm}}
    \toprule
    Corruption rate &    \\
    \midrule
    0.1 & 62.7  \\
    0.3 & 63.1   \\
    0.5 & 64.0\\
    0.7& 64.3\\
    0.9 & 63.3\\
  \bottomrule
\end{tabular}
\end{wraptable}

To further explore this hypothesis in isolation, we ran experiments \textit{corrupting} inputs $U$ (by Gaussian noise of equal mean and std) and confirm a pattern: not only does corrupting $U$ not hurt performance, but seems to \textit{boost} performance, with a $0.7$ corruption rate equivalent to our best model overall. This suggests the autoencoder quickly discerns $U_u$ is mostly noise, and adapts faster to the rich representational space constructed off of our deep encoders, resulting in a higher performance in the end. At the same time, too much corruption reveals diminishing returns: the autoencoder still requires partial observability of $U$ to bridge the latent spaces.

Our only surprise is that incorporating graph representations didn't seem to help, hence the relational information in the anime-anime suggestion graph may be more useful for recommending new animes rather than established ones, as will be demonstrated in the cold start setting.\break

\vspace{-2em}
\subsubsection{AnimeULike Cold Start}

Next, we demonstrate the improved generalization ability of DeepNaniNet in the cold start setting, particularly in handling out-of-matrix users.\hfill

%We first observe DNN to be the superior design to DN by removing the GNN component from DNN, though the difference is not as stark as in warm start. 
\begin{wraptable}{r}{8cm}
\vspace{-1.5em}
\caption{AnimeULike Cold Start. DeepNaniNet achieves superior out-of-matrix recommendation performance on anime recommendation while maintaining high in-matrix performance.}
  \begin{tabular}{p{3cm}p{2cm}p{1.5cm}}
    \toprule
    Method &  In-matrix &  Out-of-matrix  \\
    \midrule
    Popularity sort & & 13.8 \\
    Random guessing & & 10.0 \\
    DN (no UT) & 44.1 & 48.5 \\
    DN (UT) & 42.6 & 48.4 \\
    % DN (GNN + NO EIP + NO UT) & & \\
    % DN (GNN + NO EIP + HALF UT) & & \\
    % DN (NO GNN + NO EIP + NO UT)  & & \\
    % DN (NO GNN + NO EIP + HALF UT) & & \\
    \midrule
    DNN (removed GNN) & 47.9 & 52.0 \\
    DNN & \textbf{55.1} & \textbf{56.5} \\
  \bottomrule
\end{tabular}
\end{wraptable} 

 We observe superior performance of \textsc{DNN (removed GNN)} which omits the GNN representation of the item-item graph, surpassing DropoutNet. As established in the previous section, we confirm content baskets to be the superior design for cold start as well. When trained with user transform to handle out-of-matrix users, DropoutNet deteriorates on in-matrix users as a result and fails to deliver the intended boost on out-of-matrix users. DeepNaniNet, on the other hand performs better for \textbf{both} in-matrix \textit{and} out-of-matrix user representations, whereas on CiteULike it just maintains in-matrix performance. Most intriguingly, it appears on cold start items, both DropoutNet and DeepNaniNet prefer their respective out-of-matrix representations of users (user transform\footnote{It may be the case on AnimeULike, $U_u$ is basically noise already, so DropoutNet \textit{can't} overfit to $U$, causing the user transform approximation to help to an extent.} and $U_u=0$, respectively). We discuss the fascinating implication of this in a later section. Our further experiments with GNN on out-of-matrix user representations demonstrated the best results of all: 0.565 user recall@100 which highlight the true advantage of jointly learning anime representations via our item-item graph.
We believe this is due to GNN's ability to inductively generalize to new users. In particular, if new users have connections to similar items as users in the training set, the model is inductively biased towards making similar predictions for such users \cite{hamilton2017inductive}. In summary, DeepNaniNet's superior performance across both in-matrix and out-of-matrix users suggest it has constructed a richer user representational space, separate from $U$, from which it can better reconstruct relevances in the original space, in conjunction with rich anime content $\phi^V_v.$

\subsubsection{AnimeUReallyLike}
Lastly, we experiment with finetuning BERT as part of an end-to-end model. This implies end-to-end backpropagation to parameters of $\phi^V$ (and thereby $\phi^U$) via sampled content baskets.

We take a small, selected sample of 10287 (user, item) training entries and 5585 validation entries from AnimeULike that are thresholded with true relevances of >10 or <-6. We denote this small dataset of "extreme" relevances (either a user really hates or loves an anime) as \textbf{AnimeUReallyLike} and train differential language encoders for the challenging task of predicting these "extreme" preferences. Due to the resource constraints of training BERT, we leave out recall as it is not competitive and instead on preliminary findings with the validation mean square error relevance loss.

\begin{table}[H]
%   \caption{BERT validation loss, AnimeUReallyLike (New)}
  \caption{Loss curves are running averages over 50 batches to account for sampling stochasticity, trained with Adam optimizer. We found this sampling strategy achieves lower loss. Table losses are lowest achieved averages (green star in plot). For the finetuned models, we impose a 0.99 learning rate decay on BERT's unfrozen layers initialized at 1e-4.}
  \label{tab:bert_val_loss_new}
  \resizebox{\columnwidth}{!}{%
  \begin{tabular}{llllll}
    \toprule
    Models (fix drop\_p=0.75) & \makecell[l]{MSE \ref{eq:mseloss} \\ (drop\_p = 0)} & \makecell[l]{MSE \\ (drop\_p = 0.25)} & \makecell[l]{MSE \\ (drop\_p = 0.5)} & \makecell[l]{MSE \\ (drop\_p = 0.75)} & \makecell[l]{MSE \\ (drop\_p = 1)} \\
    \midrule
    bert/bert\_static, drop\_p = 0  &  10.363 & 9.947 & 10.628 & 9.689 & 10.716\\
    bert/pretrained\_bert\_static, drop\_p = 0  & 10.034 & 10.480 & 10.310 & 9.915 & 10.207\\
    bert/bert\_finetune, drop\_p = 0  & 10.304 & 10.446 & 10.134 & 9.816 & 11.029\\
    bert/pretrained\_bert\_finnetune, drop\_p = 0  & 10.595 & 9.934 & 10.519 & 9.913 & 10.183\\
  \bottomrule
\end{tabular}
}
  
\end{table}
\vspace{-2em}
\begin{figure}[H]

    \centering

    \vspace{1em}
    \resizebox{\columnwidth}{!}{%
    \includegraphics[scale=0.5]{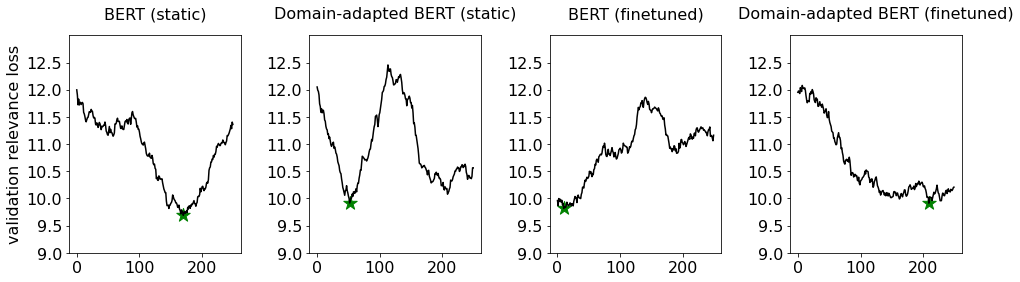}}

\end{figure}

All encoders report lower losses when drop\_p=0.75 as opposed to 0.5, attesting to the richer representations being learned via \textbf{end-to-end} backpropagation of sampled content baskets to content representations. Relatively, the loss curves suggest finetuning our domain-adapted BERT to be the superior choice for downstream training. 

While static BERT reports the single lowest value, downstream layers immediately overfit due to the language encoder's out-of-distribution representations. This can thus complicate attempts of online learning that periodically update our model to adapt to distributional shift. Meanwhile, fine-tuning BERT directly causes too large a domain shift in $\phi^V$ for tuning downstream layers (tested by independent experiments with separate learning rate schedulers for BERT layers). Instead, we advocate for finetuning our domain-adapted BERT, which consistently reports low variance and the smoothest validation loss curve across runs, as shown. Settling on the domain-adapted BERT as our anime2vec encoder, we ran experiments over all node feature encoders as well, and observed consistently lower losses with TF-IDF as the node encoder. We suspect this to be a form of feature fusion where TF-IDF features supplements the fine-tuning BERT anime2vec.

% \begin{table}
%   \caption{Transfer-finetuned BERT anime2vec: (node, edge) encoder choices, (AnimeUReallyLike)}
%   \label{tab:encoder_choices_new}
%   \begin{tabular}{lllllll}
%     \toprule
%      & N/A & BERT static & BERT finetuned & BERT transfer static & BERT transfer finetuned & TF-IDF \\
%     \midrule
%     TF-IDF & 10.718&10.677 & 10.703 & 10.721 & 10.815 & 10.760\\
%     BERT finetuned & 10.832 & 10.812 & 10.771 &10.907&10.871&10.773 \\
%     BERT transfer static & 10.769&10.617&10.850&10.816&10.847&10.757\\
%     BERT transfer finetuned &10.684&10.795&10.867&10.760&10.670&10.937\\
%     TF-IDF &  10.873& 11.174 &10.933  &10.883  &10.906  &10.999\\
%   \bottomrule
% \end{tabular}
% \end{table}
\begin{table}
  \caption{Transfer-finetuned BERT anime2vec: (node, edge) encoder choices, (AnimeUReallyLike)}
  \label{tab:encoder_choices_new}
  \resizebox{\columnwidth}{!}{%
  \begin{tabular}{llllll}
    \toprule
     (drop\_p=0.5/drop\_p=0.75) & TF-IDF & BERT static & BERT finetuned & BERT transfer static & BERT transfer finetuned  \\
    \midrule
    TF-IDF & (9.721/9.953)&(9.911/9.863) &(10.247/8.883) &(9.424/9.586)&(10.070/10.265)\\
    % BERT transfer finetuned & (9.615/9.228)&()&(10.555/10.491)&(9.883/9.961)&()\\
  \bottomrule
\end{tabular}
}
\end{table}

\begin{table}
  \caption{To test the ability to reconstruct \textit{extreme} relevances (i.e. a user's favorite show, or an item's most fervent rater), we consider the metrics of items/user MRR and users/item MRR relative to baseline encoders.}
  \label{tab:encoder_choices_new_2}
  \resizebox{\columnwidth}{!}{%
  \begin{tabular}{llllllll}
    \toprule
    Model & TF-IDF &  \makecell[l]{TF-IDF + \\GNN} & \makecell[l]{TF-IDF + \\EdgeConv} & \makecell[l]{Static \\BERT} & \makecell[l]{Fine-tuned \\BERT} & \makecell[l]{Static \\ Pre-trained BERT} & \makecell[l]{Fine-tuned \\ Pre-trained BERT} \\
    \midrule
   \makecell[l]{MRR \\(Items/User)} &  2.2e-4 (2.5e-4) & 2.3e-4&3.2e-4&2.2e-4&2.5e-4&\textbf{4.0e-4}&\textbf{5.2e-4} \\
     \midrule
    \makecell[l]{MRR \\(Users/Item)} & 4.1e-4 (1.6e-4)&3.6e-4&8.7e-4&\textbf{11.5e-4}&\textbf{8.7e-4}&2.6e-4&2.3e-4 \\
  \bottomrule
\end{tabular}
}
\end{table}

% \begin{table}[H]
% \begin{tabular}{|p{1.5cm}||p{1.5cm}|p{1.5cm}|p{1.5cm}|p{1.5cm}|p{1.5cm}|p{1.5cm}|p{1.5cm}|}
% \hline
% Model & TF-IDF &  TF-IDF + GNN & TF-IDF + EdgeConv & Static BERT & Fine-tuned BERT & Static Pre-trained BERT & Fine-tuned Pre-trained BERT\\ \hline
%  MRR (Items/User) &  2.2e-4 (2.5e-4) & 2.3e-4&3.2e-4&2.2e-4&2.5e-4&\textbf{4.0e-4}&\textbf{5.2e-4}\\
%  MRR (Users/Item) & 4.1e-4 (1.6e-4)&3.6e-4&8.7e-4&\textbf{11.5e-4}&\textbf{8.7e-4}&2.6e-4&2.3e-4\\
%  \hline
% \end{tabular}
% \end{table}

% \begin{table}
%   \caption{DeepNaniNet, AnimeUReallyLike, BERT transfer finetuned anime2vec (New)}
%   \label{tab:deep_naninet_loss}
%   \begin{tabular}{lll}
%     \toprule
%     Models (node, edge) & Loss (drop\_p = 0.5) & Loss (drop\_p = 0.75) \\
%     \midrule
%     ('tf-idf', 'bert-finetuned') & 10.247 & 8.883 \\
%     ('tf-idf', 'tf-idf')& 9.721 & 9.953\\
%     ('bert-finetuned', 'tf-idf') & 10.097 & 10.310 \\
%     ('bert-transfer-finetuned', 'tf-idf') & 9.615 & 9.228 \\
%     ('tf-idf', 'bert-transfer-finetuned')& 10.070 & 10.265 \\
%     ('tf-idf', 'bert-transfer-static') &  9.424 & 9.586 \\
%     ('bert-transfer-static', 'tf-idf') &  10.339 & 9.983 \\
%     ('bert-transfer-finetuned', 'bert-transfer-static') &  11.429 & \\
%     ('bert-static', 'tf-idf') & 9.727 & \\
%     ('tf-idf', 'bert-static') & 9.911 & \\
%   \bottomrule
% \end{tabular}
% \end{table}

The expected result of \textit{random guessing} is in parentheses. Despite seeing only a small subset (4592 positive, 6235 negative entries), these models fare multiples better than random chance. Transferring BERT in-domain is advantageous for retrieving a user's top show due to pretraining on a diverse set of high quality user reviews. Moreover, the BERT models outshine TF-IDF in at least one MRR department. We suspect that while TF-IDF is a better comprehensive encoder in non-semantic domains (i.e. CiteULike) due to high correlation in features, it struggles to find a decision boundary in semantic domains where most tokens appear in both extremely positive and extremely negative cases. Thus, a second case for deep language encoders is their complex non-linear decision boundaries that excel in semantic settings. We discuss these advantages further in the section below.

\section{Discussion}
%Your report should include \textit{qualitative evaluation}. That is, try to understand your system (e.g. how it works, when it succeeds and when it fails) by inspecting key characteristics or outputs of your model.
\subsection{Qualitative Analysis}

In our \href{http://otakuroll.net}{app}, the user experience using \textbf{DeepNaniNet} with WMF and top-K CF feels drastically different. Users have complained WMF only returns what's already popular. While top-K CF's recommendations feel a bit more intentional (as adopted by the best attempts to build anime recommendation engines like \href{https://www.anirec.net/}{here}), it achieves little more than recommendations you may get after consulting multiple friends. Our approach consistently captures underrated anime (ones that fly under the radar but when you look into its synopsis, is thematic), making its high recall numbers all the more impressive. As an example, our prior prototype (which recommended exclusively from the top ~250 shows for more traction) produced the following when querying with two highly regarded shounens. 

["Fullmetal Alchemist Brotherhood" (FMAB)\footnote{FMAB is rated \#1 on MAL all-time and AoT is trending \#1 in popularity.}, "Shingeky No Kyojin" (AoT/Attack on Titan)] $\Rightarrow$  \begin{itemize}
    \item WMF: 1) Haikyuu, 2) Hunter x Hunter \footnote{Both Haikyuu and Hunter x Hunter are among the most popular shounen.}
    \item Top-K: 1) Mushishi, 2) Stein's Gate, 3) Haikyuu, 4) Hunter x Hunter \footnote{Mushishi is irrelevant in content but has \textit{a lot} of item-item edges (preferred by CF) hence a diverse rater base. Stein's Gate is highly regarded but not action/adventure shounen, again demonstrating the popularity bias.}
    \item DeepNaniNet: 1) Hajime no Ippo \footnote{Highly underrated shounen!}, 2) Nana\footnote{A lesser known slice of life, has a synopsis that says two girls travelling together in search of one girl's boyfriend (note: FMAB's synopsis talks about two brothers travelling together in search of the philosopher's stone!)}, 3) Gintama \footnote{A bit of an offbeat satirical shounen that pokes fun at other shounens!}
\end{itemize}

We make the following generalizations: WMF is biased towards popular anime. Top-K is biased towards highly rated anime with lots of item-item recommendations, while \textbf{DeepNaniNet} seems capable of \textit{actually reading} information on other shows and outputting thematically similar (and often underrated) shows. Perhaps the biggest advantage of DeepNaniNet comes from the fact it actually represents all available information word-for-word to make recommendations. This avoids content popularity biases and consistently outputs underrated shows. However, it also comes with high sensitivity to strong sentiment and word-level mentions, esp. on AnimeUReallyLike. An example is when running the AnimeUReallyLike model on the same query set ['Fullmetal Alchemist Brotherhood', 'Shingeky No Kyojin'] (FMAB and AoT), two of the highest rated/regarded shounen series of all time.
We observe:
\begin{itemize}
    \item TF-IDF GNN: A Farewell to Arms\footnote{A Farewell to Arms: a story of "power suit-wearin' men tasked with disarming automatic tanks in a post-apocalyptic Tokyo"}, Armored Trooper Votoms\footnote{Armored Trooper Votoms: set in "a century of bloodshed between warring star systems... flames of war..." where "a special forces powered-armor pilot is suddenly transferred into a unit engaged in a secert and highly illegal mission to steal military secrets..." (you get the idea)}, Hotori\footnote{Hotori: At the Personality Plant, robots are being built and slowly outfitted with the artificial memories of real people." The main character, Suzu, "is one such robot."}, Sailor Moon SuperS the Movie
    \item TF-IDF GINE: Armored Trooper Votoms\footnote{Armored Trooper Votoms: set in a city "built form the labors of mechanical beasts... with incredible destructive power as a new type of advanced weaponry"}, A Farewell to Arms, Slayers, Patlabor: The Movie
    \item Finetuned BERT: Royal Space Force\footnote{Royal Space Force: Protagonist is part of the country's space force, who embark on a mission to redeem humanity by restoring its strength}, Desert Punk, Ginga Eiyuu Densetsu\footnote{Ginga Eiyuu Densetsu: About a coup staged by the National Salvation Military Council under the direction of the Galactic Empire, happening during civil wars in both the Alliance and the Empire}, Mobile Sui Gundam\footnote{Mobile Sui Gundam: About a space immigrant who joins the League Militaire, a militia frustrated with their empire's cruelty, who fights to bring an end to the Zanscare Empire's reign}, Azur Lane\footnote{Azur Lane: pits "a divided humanity" which "stood in complete solidarity" against "an alien force with an arsenal far surpassing the limits of current technology"; with countries joining forces, "paving the way for the improvement of modern warfare"... during "neverending conflict within humankind" (basically if FMAB and AoT had a baby... this would be it!)}
    \item Fine-tuned transferred BERT: .hack//Sign, .hack//Quantum, Fate/Zero, .hack//Liminality, .hack//Gift
\end{itemize}

We observe GNN and GINE both converge on a similar set of results that all common themes of military, humanity's war, revolution, etc. Reading FMAB and AoT's descriptions, we see these themes present in their synopses: "military allies, colonel, lieutenant, nationwide conspiracy, state, law" and "humanity, extinction, defensive barriers, fight for survival, Survey Corps, military unit, brutal war, walls". Fascinatingly, we observe plot-level similarities with FMAB and Hotori. FMAB is about two brothers who lost parts of their physical bodies. "It is the hope that they would both eventually return to their original bodies..." whereas Hotori has this identical element.

TF-IDF seems capable of capturing similarities from narrow dimensions (such as shallow mentions of military themes), even if those dimensions are not the central themes of either FMAB or AoT. 

Fine-tuning BERT, meanwhile, \textbf{captures similarity across more complex dimensions}. The shows are more diverse in plot while rooted in common themes across militant conflict, failure of government, humanity, and revolution.

At the same time, this turned out as an adversial example for the transferred BERT encoder. We discovered the synopses and reviews for shows within the ".hack" franchise (aptly named) repetitively refer to the series as a whole, making the language encoder sensitive to its mentions as opposed to learning each show's distinct content, resulting in the four of them to be recommended together. Looking past that, Fate/Zero is a \textit{very} good recommendation\footnote{Fate/Zero: not only thematically similar (war, battle royale, etc.) but is regarded as a crossover between both FMAB and AoT: exploration of deep themes and unapologetic cruelty}. Nonetheless, this adversarial example poses a concern of using deep language encoders four our system. We are motivated to pursue, as a direction of future study, the effect on recommendations due to the encoder's sensitivity to word mentions or extreme forms of sentiment.

\subsection{Applications}
Our system is extremely practical for inference: only $\phi^V, \hat{U},\hat{V}$ needs to be cached (only the latter two for in-matrix users) and an additional step of computing $\hat{U}$ off of $\phi^V$ for guest users with the option of approximating $\hat{U_{guest}}\approx \frac{1}{|V({u_{guest}}|}\sum_{v\in V({u_{guest}})}\hat{V_v}$. This enables parallelism for faster retrieval. This is a huge win for SaaS recommendation services bootstrapping off minimal user data. We envision the flourishing of open-source representations for popular items across popular culture and media, enabling more niche services to experiment with ways of content basket design for satisfying more domain-specific tastes. Guest users can then experiment and enjoy high-quality recommendations with these services without fear of being mined of personal data.

For large companies, this can avoid many of the privacy concerns and technical pains of storing, managing, and exploiting a customer's entire lifecycle on the application. As companies adopt more content-based recommendation systems, we believe \textit{latent} modelling of $\phi^U$ conditional on context (i.e. $\phi^U_u=\sum_{v\in V(u)}f(v|u,c)\phi^V_v$ weighted content baskets) can design more intentional recommendations (as in Spotify's user explanatory framework) dependent on a user's context $c$ (i.e. "looking for fantasy" or "pumped up"). 

Due to our strong performance on cold start, we believe content creators and advertising channels can use services built off this model to test potential audience traction with new types of content, whose deep representations avoid the negative feedback loop of collaborative-filtering approaches.

\section{Conclusion}
%Summarize the main findings of your project, and what you have learnt. Highlight your achievements, and note the primary limitations of your work. If you like, you can describe avenues for future work.
We introduced DeepNaniNet, a neural recommender system framework for reconstructing user-item preferences via rich content encodings, and our techniques for better cold start generalization. We replicated DropoutNet's SOTA cold start results on CiteULike, where our model maintains equally strong performance across the out vs. in-matrix users, hence outperforming DropoutNet in the realistic real-world setting where 50\% of users are guest users. We introduced AnimeULike, a dataset rich in content but sparse in preferences, and demonstrated strong performance on both warm and cold start - notably a ~7-fold improvement over the WMF baseline - including further experiments revealing rich properties of a denoising autoencoder. We demonstrated further gains in generalization to cold start items via jointly learnt graph representations. Finally, we made the case for deep, differentiable language encoders and feasibility for end-to-end training. Lastly, we close with our motivation that started it all: to deliver more meaningful, personalized, and engaging content for users (whether old, new or guest) without compromising our principles for user privacy.
% \begin{acks}
% \end{acks}

\bibliographystyle{ACM-Reference-Format}
\bibliography{recsys-manuscript-bib}


\begin{thebibliography}{26}

%%% the Web address.

\ifx \showCODEN    \undefined \def \showCODEN     #1{\unskip}     \fi
\ifx \showDOI      \undefined \def \showDOI       #1{#1}\fi
\ifx \showISBNx    \undefined \def \showISBNx     #1{\unskip}     \fi
\ifx \showISBNxiii \undefined \def \showISBNxiii  #1{\unskip}     \fi
\ifx \showISSN     \undefined \def \showISSN      #1{\unskip}     \fi
\ifx \showLCCN     \undefined \def \showLCCN      #1{\unskip}     \fi
\ifx \shownote     \undefined \def \shownote      #1{#1}          \fi
\ifx \showarticletitle \undefined \def \showarticletitle #1{#1}   \fi
\ifx \showURL      \undefined \def \showURL       {\relax}        \fi
% invisible to TeX
\providecommand\bibfield[2]{#2}
\providecommand\bibinfo[2]{#2}
\providecommand\natexlab[1]{#1}
\providecommand\showeprint[2][]{arXiv:#2}

\bibitem[\protect\citeauthoryear{Ahmadian, Afsharchi, and Meghdadi}{Ahmadian
  et~al\mbox{.}}{2019}]%
        {ahmadian2019novel}
\bibfield{author}{\bibinfo{person}{Sajad Ahmadian}, \bibinfo{person}{Mohsen
  Afsharchi}, {and} \bibinfo{person}{Majid Meghdadi}.}
  \bibinfo{year}{2019}\natexlab{}.
\newblock \showarticletitle{A novel approach based on multi-view reliability
  measures to alleviate data sparsity in recommender systems}.
\newblock \bibinfo{journal}{\emph{Multimedia Tools and Applications}}
  \bibinfo{volume}{78}, \bibinfo{number}{13} (\bibinfo{year}{2019}),
  \bibinfo{pages}{17763--17798}.
\newblock


\bibitem[\protect\citeauthoryear{Bernardi, Kamps, Kiseleva, and
  M{\"u}ller}{Bernardi et~al\mbox{.}}{2015}]%
        {bernardi2015continuous}
\bibfield{author}{\bibinfo{person}{Lucas Bernardi}, \bibinfo{person}{Jaap
  Kamps}, \bibinfo{person}{Julia Kiseleva}, {and} \bibinfo{person}{Melanie~JI
  M{\"u}ller}.} \bibinfo{year}{2015}\natexlab{}.
\newblock \showarticletitle{The continuous cold start problem in e-commerce
  recommender systems}.
\newblock \bibinfo{journal}{\emph{arXiv preprint arXiv:1508.01177}}
  (\bibinfo{year}{2015}).
\newblock


\bibitem[\protect\citeauthoryear{Bobadilla, Ortega, Hernando, and
  Gutiérrez}{Bobadilla et~al\mbox{.}}{2013}]%
        {recsurvey}
\bibfield{author}{\bibinfo{person}{J. Bobadilla}, \bibinfo{person}{F. Ortega},
  \bibinfo{person}{A. Hernando}, {and} \bibinfo{person}{A. Gutiérrez}.}
  \bibinfo{year}{2013}\natexlab{}.
\newblock \showarticletitle{Recommender systems survey}.
\newblock \bibinfo{journal}{\emph{Knowledge-Based Systems}}
  \bibinfo{volume}{46} (\bibinfo{year}{2013}), \bibinfo{pages}{109--132}.
\newblock
\showISSN{0950-7051}
\urldef\tempurl%
\url{https://doi.org/10.1016/j.knosys.2013.03.012}
\showDOI{\tempurl}


\bibitem[\protect\citeauthoryear{Bogers and Van~den Bosch}{Bogers and Van~den
  Bosch}{2008}]%
        {bogers2008recommending}
\bibfield{author}{\bibinfo{person}{Toine Bogers} {and} \bibinfo{person}{Antal
  Van~den Bosch}.} \bibinfo{year}{2008}\natexlab{}.
\newblock \showarticletitle{Recommending scientific articles using citeulike}.
  In \bibinfo{booktitle}{\emph{Proceedings of the 2008 ACM conference on
  Recommender systems}}. \bibinfo{pages}{287--290}.
\newblock


\bibitem[\protect\citeauthoryear{Chee, Han, and Wang}{Chee
  et~al\mbox{.}}{2001}]%
        {chee2001rectree}
\bibfield{author}{\bibinfo{person}{Sonny Han~Seng Chee},
  \bibinfo{person}{Jiawei Han}, {and} \bibinfo{person}{Ke Wang}.}
  \bibinfo{year}{2001}\natexlab{}.
\newblock \showarticletitle{Rectree: An efficient collaborative filtering
  method}. In \bibinfo{booktitle}{\emph{International Conference on Data
  Warehousing and Knowledge Discovery}}. Springer, \bibinfo{pages}{141--151}.
\newblock


\bibitem[\protect\citeauthoryear{Devlin, Chang, Lee, and Toutanova}{Devlin
  et~al\mbox{.}}{2018}]%
        {devlin2018bert}
\bibfield{author}{\bibinfo{person}{Jacob Devlin}, \bibinfo{person}{Ming-Wei
  Chang}, \bibinfo{person}{Kenton Lee}, {and} \bibinfo{person}{Kristina
  Toutanova}.} \bibinfo{year}{2018}\natexlab{}.
\newblock \showarticletitle{Bert: Pre-training of deep bidirectional
  transformers for language understanding}.
\newblock \bibinfo{journal}{\emph{arXiv preprint arXiv:1810.04805}}
  (\bibinfo{year}{2018}).
\newblock


\bibitem[\protect\citeauthoryear{Eksombatchai, Jindal, Liu, Liu, Sharma,
  Sugnet, Ulrich, and Leskovec}{Eksombatchai et~al\mbox{.}}{2018}]%
        {10.1145/3178876.3186183}
\bibfield{author}{\bibinfo{person}{Chantat Eksombatchai},
  \bibinfo{person}{Pranav Jindal}, \bibinfo{person}{Jerry~Zitao Liu},
  \bibinfo{person}{Yuchen Liu}, \bibinfo{person}{Rahul Sharma},
  \bibinfo{person}{Charles Sugnet}, \bibinfo{person}{Mark Ulrich}, {and}
  \bibinfo{person}{Jure Leskovec}.} \bibinfo{year}{2018}\natexlab{}.
\newblock \showarticletitle{Pixie: A system for recommending 3+ billion items
  to 200+ million users in real-time}. In \bibinfo{booktitle}{\emph{Proceedings
  of the 2018 world wide web conference}}. \bibinfo{pages}{1775--1784}.
\newblock


\bibitem[\protect\citeauthoryear{Gopalan, Hofman, and Blei}{Gopalan
  et~al\mbox{.}}{2013}]%
        {ctpr}
\bibfield{author}{\bibinfo{person}{Prem Gopalan}, \bibinfo{person}{Jake~M
  Hofman}, {and} \bibinfo{person}{David~M Blei}.}
  \bibinfo{year}{2013}\natexlab{}.
\newblock \showarticletitle{Scalable recommendation with poisson
  factorization}.
\newblock \bibinfo{journal}{\emph{arXiv preprint arXiv:1311.1704}}
  (\bibinfo{year}{2013}).
\newblock


\bibitem[\protect\citeauthoryear{Hamilton, Ying, and Leskovec}{Hamilton
  et~al\mbox{.}}{2017}]%
        {hamilton2017inductive}
\bibfield{author}{\bibinfo{person}{William~L. Hamilton}, \bibinfo{person}{Rex
  Ying}, {and} \bibinfo{person}{Jure Leskovec}.}
  \bibinfo{year}{2017}\natexlab{}.
\newblock \showarticletitle{Inductive Representation Learning on Large Graphs}.
  In \bibinfo{booktitle}{\emph{NIPS}}.
\newblock


\bibitem[\protect\citeauthoryear{Hofmann}{Hofmann}{2004}]%
        {10.1145/963770.963774}
\bibfield{author}{\bibinfo{person}{Thomas Hofmann}.}
  \bibinfo{year}{2004}\natexlab{}.
\newblock \showarticletitle{Latent semantic models for collaborative
  filtering}.
\newblock \bibinfo{journal}{\emph{ACM Transactions on Information Systems
  (TOIS)}} \bibinfo{volume}{22}, \bibinfo{number}{1} (\bibinfo{year}{2004}),
  \bibinfo{pages}{89--115}.
\newblock


\bibitem[\protect\citeauthoryear{Hu*, Liu*, Gomes, Zitnik, Liang, Pande, and
  Leskovec}{Hu* et~al\mbox{.}}{2020}]%
        {Hu*2020Strategies}
\bibfield{author}{\bibinfo{person}{Weihua Hu*}, \bibinfo{person}{Bowen Liu*},
  \bibinfo{person}{Joseph Gomes}, \bibinfo{person}{Marinka Zitnik},
  \bibinfo{person}{Percy Liang}, \bibinfo{person}{Vijay Pande}, {and}
  \bibinfo{person}{Jure Leskovec}.} \bibinfo{year}{2020}\natexlab{}.
\newblock \showarticletitle{Strategies for Pre-training Graph Neural Networks}.
  In \bibinfo{booktitle}{\emph{International Conference on Learning
  Representations}}.
\newblock
\urldef\tempurl%
\url{https://openreview.net/forum?id=HJlWWJSFDH}
\showURL{%
\tempurl}


\bibitem[\protect\citeauthoryear{Hu, Koren, and Volinsky}{Hu
  et~al\mbox{.}}{2008}]%
        {hu2008collaborative}
\bibfield{author}{\bibinfo{person}{Yifan Hu}, \bibinfo{person}{Yehuda Koren},
  {and} \bibinfo{person}{Chris Volinsky}.} \bibinfo{year}{2008}\natexlab{}.
\newblock \showarticletitle{Collaborative filtering for implicit feedback
  datasets}. In \bibinfo{booktitle}{\emph{2008 Eighth IEEE International
  Conference on Data Mining}}. Ieee, \bibinfo{pages}{263--272}.
\newblock


\bibitem[\protect\citeauthoryear{Kipf and Welling}{Kipf and Welling}{2017}]%
        {kipf2017semi}
\bibfield{author}{\bibinfo{person}{Thomas~N. Kipf} {and} \bibinfo{person}{Max
  Welling}.} \bibinfo{year}{2017}\natexlab{}.
\newblock \showarticletitle{Semi-Supervised Classification with Graph
  Convolutional Networks}. In \bibinfo{booktitle}{\emph{International
  Conference on Learning Representations (ICLR)}}.
\newblock


\bibitem[\protect\citeauthoryear{Lika, Kolomvatsos, and Hadjiefthymiades}{Lika
  et~al\mbox{.}}{2014}]%
        {lika2014facing}
\bibfield{author}{\bibinfo{person}{Blerina Lika}, \bibinfo{person}{Kostas
  Kolomvatsos}, {and} \bibinfo{person}{Stathes Hadjiefthymiades}.}
  \bibinfo{year}{2014}\natexlab{}.
\newblock \showarticletitle{Facing the cold start problem in recommender
  systems}.
\newblock \bibinfo{journal}{\emph{Expert Systems with Applications}}
  \bibinfo{volume}{41}, \bibinfo{number}{4} (\bibinfo{year}{2014}),
  \bibinfo{pages}{2065--2073}.
\newblock


\bibitem[\protect\citeauthoryear{Singh}{Singh}{2020}]%
        {singh2020scalability}
\bibfield{author}{\bibinfo{person}{Monika Singh}.}
  \bibinfo{year}{2020}\natexlab{}.
\newblock \showarticletitle{Scalability and sparsity issues in recommender
  datasets: a survey}.
\newblock \bibinfo{journal}{\emph{Knowledge and Information Systems}}
  \bibinfo{volume}{62}, \bibinfo{number}{1} (\bibinfo{year}{2020}),
  \bibinfo{pages}{1--43}.
\newblock


\bibitem[\protect\citeauthoryear{Su and Khoshgoftaar}{Su and
  Khoshgoftaar}{2006}]%
        {su2006collaborative}
\bibfield{author}{\bibinfo{person}{Xiaoyuan Su} {and} \bibinfo{person}{Taghi~M
  Khoshgoftaar}.} \bibinfo{year}{2006}\natexlab{}.
\newblock \showarticletitle{Collaborative filtering for multi-class data using
  belief nets algorithms}. In \bibinfo{booktitle}{\emph{2006 18th IEEE
  international conference on Tools with Artificial Intelligence (ICTAI'06)}}.
  IEEE, \bibinfo{pages}{497--504}.
\newblock


\bibitem[\protect\citeauthoryear{Van Den~Oord, Dieleman, and Schrauwen}{Van
  Den~Oord et~al\mbox{.}}{2013}]%
        {deepmusic}
\bibfield{author}{\bibinfo{person}{A{\"a}ron Van Den~Oord},
  \bibinfo{person}{Sander Dieleman}, {and} \bibinfo{person}{Benjamin
  Schrauwen}.} \bibinfo{year}{2013}\natexlab{}.
\newblock \showarticletitle{Deep content-based music recommendation}. In
  \bibinfo{booktitle}{\emph{Neural Information Processing Systems Conference
  (NIPS 2013)}}, Vol.~\bibinfo{volume}{26}. Neural Information Processing
  Systems Foundation (NIPS).
\newblock


\bibitem[\protect\citeauthoryear{Volkovs, Yu, and Poutanen}{Volkovs
  et~al\mbox{.}}{2017}]%
        {dropout}
\bibfield{author}{\bibinfo{person}{Maksims Volkovs}, \bibinfo{person}{Guang~Wei
  Yu}, {and} \bibinfo{person}{Tomi Poutanen}.} \bibinfo{year}{2017}\natexlab{}.
\newblock \showarticletitle{DropoutNet: Addressing Cold Start in Recommender
  Systems.}. In \bibinfo{booktitle}{\emph{NIPS}}. \bibinfo{pages}{4957--4966}.
\newblock


\bibitem[\protect\citeauthoryear{Wang and Blei}{Wang and Blei}{2011}]%
        {ctr}
\bibfield{author}{\bibinfo{person}{Chong Wang} {and} \bibinfo{person}{David~M
  Blei}.} \bibinfo{year}{2011}\natexlab{}.
\newblock \showarticletitle{Collaborative topic modeling for recommending
  scientific articles}. In \bibinfo{booktitle}{\emph{Proceedings of the 17th
  ACM SIGKDD international conference on Knowledge discovery and data mining}}.
  \bibinfo{pages}{448--456}.
\newblock


\bibitem[\protect\citeauthoryear{Wang, Wang, and Yeung}{Wang
  et~al\mbox{.}}{2015}]%
        {cdl}
\bibfield{author}{\bibinfo{person}{Hao Wang}, \bibinfo{person}{Naiyan Wang},
  {and} \bibinfo{person}{Dit-Yan Yeung}.} \bibinfo{year}{2015}\natexlab{}.
\newblock \showarticletitle{Collaborative deep learning for recommender
  systems}. In \bibinfo{booktitle}{\emph{Proceedings of the 21th ACM SIGKDD
  international conference on knowledge discovery and data mining}}.
  \bibinfo{pages}{1235--1244}.
\newblock


\bibitem[\protect\citeauthoryear{Wolf, Debut, Sanh, Chaumond, Delangue, Moi,
  Cistac, Rault, Louf, Funtowicz, et~al\mbox{.}}{Wolf et~al\mbox{.}}{2019}]%
        {wolf2019huggingface}
\bibfield{author}{\bibinfo{person}{Thomas Wolf}, \bibinfo{person}{Lysandre
  Debut}, \bibinfo{person}{Victor Sanh}, \bibinfo{person}{Julien Chaumond},
  \bibinfo{person}{Clement Delangue}, \bibinfo{person}{Anthony Moi},
  \bibinfo{person}{Pierric Cistac}, \bibinfo{person}{Tim Rault},
  \bibinfo{person}{R{\'e}mi Louf}, \bibinfo{person}{Morgan Funtowicz},
  {et~al\mbox{.}}} \bibinfo{year}{2019}\natexlab{}.
\newblock \showarticletitle{HuggingFace's Transformers: State-of-the-art
  natural language processing}.
\newblock \bibinfo{journal}{\emph{arXiv preprint arXiv:1910.03771}}
  (\bibinfo{year}{2019}).
\newblock


\bibitem[\protect\citeauthoryear{Wu, Ahmed, Beutel, Smola, and Jing}{Wu
  et~al\mbox{.}}{2017}]%
        {rnn}
\bibfield{author}{\bibinfo{person}{Chao-Yuan Wu}, \bibinfo{person}{Amr Ahmed},
  \bibinfo{person}{Alex Beutel}, \bibinfo{person}{Alexander~J Smola}, {and}
  \bibinfo{person}{How Jing}.} \bibinfo{year}{2017}\natexlab{}.
\newblock \showarticletitle{Recurrent recommender networks}. In
  \bibinfo{booktitle}{\emph{Proceedings of the tenth ACM international
  conference on web search and data mining}}. \bibinfo{pages}{495--503}.
\newblock


\bibitem[\protect\citeauthoryear{Xue, Dai, Zhang, Huang, and Chen}{Xue
  et~al\mbox{.}}{2017}]%
        {xue2017deep}
\bibfield{author}{\bibinfo{person}{Hong-Jian Xue}, \bibinfo{person}{Xinyu Dai},
  \bibinfo{person}{Jianbing Zhang}, \bibinfo{person}{Shujian Huang}, {and}
  \bibinfo{person}{Jiajun Chen}.} \bibinfo{year}{2017}\natexlab{}.
\newblock \showarticletitle{Deep Matrix Factorization Models for Recommender
  Systems.}. In \bibinfo{booktitle}{\emph{IJCAI}}, Vol.~\bibinfo{volume}{17}.
  Melbourne, Australia, \bibinfo{pages}{3203--3209}.
\newblock


\bibitem[\protect\citeauthoryear{Ying, He, Chen, Eksombatchai, Hamilton, and
  Leskovec}{Ying et~al\mbox{.}}{2018}]%
        {ying2018graph}
\bibfield{author}{\bibinfo{person}{Rex Ying}, \bibinfo{person}{Ruining He},
  \bibinfo{person}{Kaifeng Chen}, \bibinfo{person}{Pong Eksombatchai},
  \bibinfo{person}{William~L Hamilton}, {and} \bibinfo{person}{Jure Leskovec}.}
  \bibinfo{year}{2018}\natexlab{}.
\newblock \showarticletitle{Graph convolutional neural networks for web-scale
  recommender systems}. In \bibinfo{booktitle}{\emph{Proceedings of the 24th
  ACM SIGKDD International Conference on Knowledge Discovery \& Data Mining}}.
  \bibinfo{pages}{974--983}.
\newblock


\bibitem[\protect\citeauthoryear{Zhang, Shi, Zhao, and King}{Zhang
  et~al\mbox{.}}{2019}]%
        {zhang2019star}
\bibfield{author}{\bibinfo{person}{Jiani Zhang}, \bibinfo{person}{Xingjian
  Shi}, \bibinfo{person}{Shenglin Zhao}, {and} \bibinfo{person}{Irwin King}.}
  \bibinfo{year}{2019}\natexlab{}.
\newblock \showarticletitle{STAR-GCN: stacked and reconstructed graph
  convolutional networks for recommender systems}.
\newblock \bibinfo{journal}{\emph{arXiv preprint arXiv:1905.13129}}.
\newblock


\bibitem[\protect\citeauthoryear{{Zhang}, {Liu}, {Zhang}, and {Zhou}}{{Zhang}
  et~al\mbox{.}}{2010}]%
        {socialtag}
\bibfield{author}{\bibinfo{person}{Zi-Ke {Zhang}}, \bibinfo{person}{Chuang
  {Liu}}, \bibinfo{person}{Yi-Cheng {Zhang}}, {and} \bibinfo{person}{Tao
  {Zhou}}.} \bibinfo{year}{2010}\natexlab{}.
\newblock \showarticletitle{{Solving the cold-start problem in recommender
  systems with social tags}}.
\newblock \bibinfo{journal}{\emph{EPL (Europhysics Letters)}}
  \bibinfo{volume}{92}, \bibinfo{number}{2} (\bibinfo{date}{Oct.}
  \bibinfo{year}{2010}), \bibinfo{pages}{28002}.
\newblock
\urldef\tempurl%
\url{https://doi.org/10.1209/0295-5075/92/28002}
\showDOI{\tempurl}
\showeprint[arxiv]{1004.3732}~[cs.IR]


\end{thebibliography}

\appendix

\end{document}